\documentclass[aip,pop,reprint]{revtex4-1}

\usepackage{graphicx}
\usepackage{dcolumn}
\usepackage{bm}

\usepackage[utf8]{inputenc}
\usepackage[T1]{fontenc}
\usepackage{mathptmx}
\usepackage{etoolbox}

\begin{document}


\title{Density-gradient-driven drift waves in the solar corona} 



\author{M. Brchnelova}
\email{michaela.brchnelova@kuleuven.be}
\affiliation{Centre for Mathematical Plasma Astrophysics, KU Leuven, Celestijnenlaan 200B, 3001, Leuven}

\author{MJ Pueschel}
\affiliation{Dutch Institute for Fundamental Energy Research, 5612 AJ Eindhoven, The Netherlands}
\affiliation{Eindhoven University of Technology, 5600 MB Eindhoven, The Netherlands}
\affiliation{Department of Physics \& Astronomy, Ruhr-Universit{\"a}t Bochum, D-44780 Bochum, Germany}

\author{S. Poedts}
\altaffiliation[Also at ]{Centre for Mathematical Plasma Astrophysics, KU Leuven, Celestijnenlaan 200B, 3001, Leuven}
\affiliation{Institute of Physics, University of Maria Curie-Sk{\l}odowska, ul.\ Radziszewskiego 10, 20-031 Lublin, Poland }

\date{\today}

\begin{abstract}
It has been suggested that under solar coronal conditions, drift waves may contribute to coronal heating. Specific properties of the drift waves to be expected in the solar corona have, however, not yet been determined using more advanced numerical models. We investigate the linear properties of density-gradient-driven drift waves in the solar coronal plasma using gyrokinetic ion-electron simulations with the gyrokinetic code \textsc{Gene}, solving the Vlasov-Maxwell equations in five dimensions assuming a simple slab geometry. We determine the frequencies and growth rates of the coronal density gradient-driven drift waves with changing plasma parameters, such as the electron $\beta$, the density gradient, the magnetic shear and additional temperature gradients. To investigate the influence of the finite Larmor radius effect on the growth and structure of the modes, we also compare the gyrokinetic simulation results to those obtained from drift-kinetics. In {most} of the investigated conditions, the drift wave has positive growth rates that increase with increasing density gradient and decreasing $\beta$. In the case of increasing magnetic shear, we find that from a certain point, the growth rate reaches a plateau. Depending on the considered reference environment, the frequencies and growth rates of these waves lie on the order of $0.1\;$mHz to $1\;$Hz. These values correspond to the observed solar wind density fluctuations near the Sun detected by WISPR, currently of unexplained origin. As a next step, nonlinear simulations are required to determine the expected fluctuation amplitudes and the plasma heating resulting from this mechanism.
\end{abstract}


\maketitle 

\section{Drift waves in the solar corona}
\label{sec:introduction}

A common approach to describing the behaviour of plasma in the solar corona is to rely on the ideal magnetohydrodynamic (MHD) equations. These equations, which govern the dynamics of a single-fluid  plasma with zero resistivity, can relatively accurately predict global plasma characteristics, such as some of the plasma flow phenomena and equilibrium conditions, and resolve specific waves potentially present in the plasma, including magnetosonic and Alfvén waves. This model, however, cannot resolve different dynamics of the ions and the electrons, and even in the cases of more advanced multi-fluid MHD modelling, MHD approaches still cannot resolve the associated kinetic effects.

The extra degree of freedom that results from the separate ion and electron motions in the plasma also adds supplementary mechanisms that can destabilise the plasma, in addition to the modes already known from MHD theory. One such set of phenomena, called drift waves (DWs), prescribes that there is always a potential for instability if the plasma has a spatial gradient in the distribution function of its particles, which is a phenomenon that cannot be predicted or modelled by the MHD theory alone.

   \begin{figure*}
   \includegraphics[width=\textwidth]{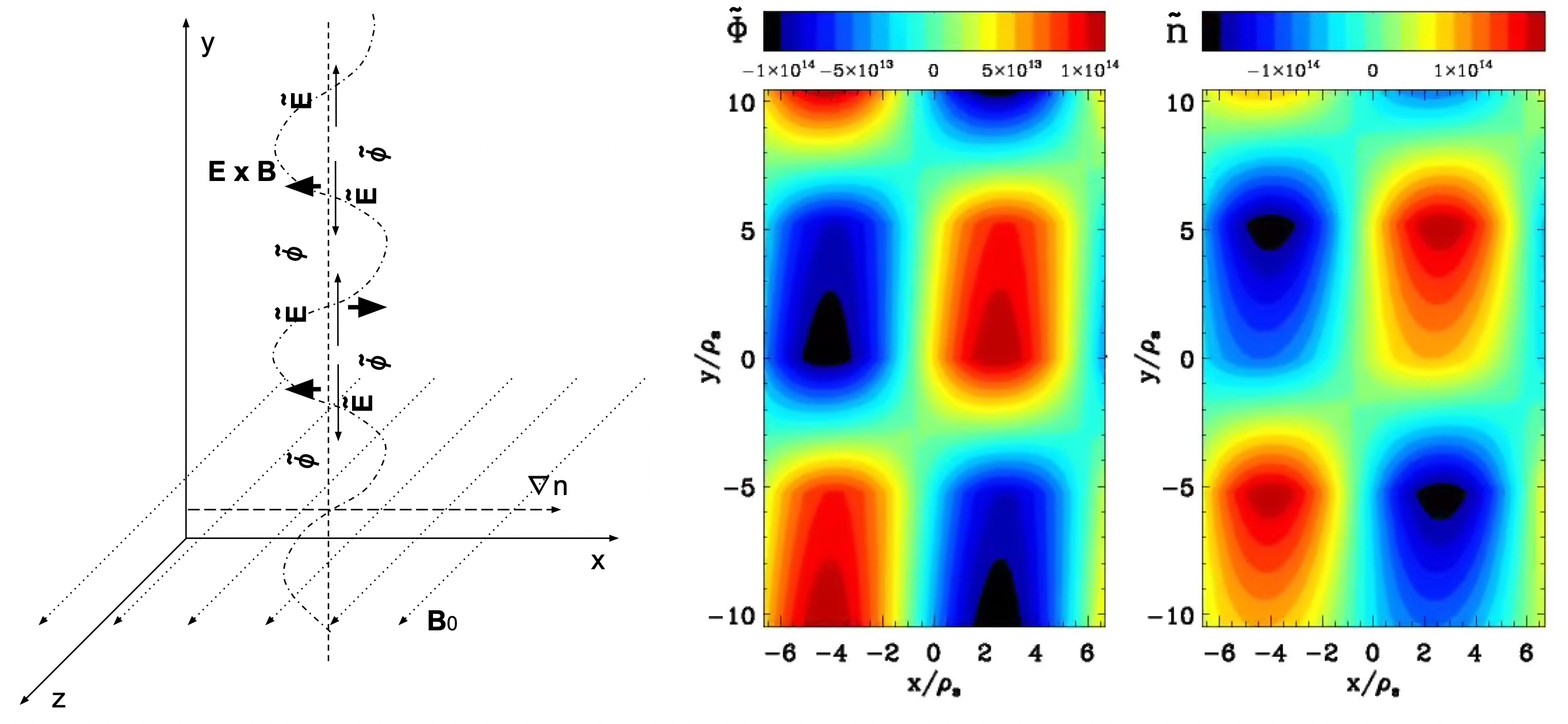}
   \caption{On the left is visualised a schematic of the drift wave mechanism as explained in the text. Shown on the right are example contours of a $k_y = 0.3$ drift-wave instability in terms of the perturbed electrostatic potential and electron number density as functions of the coordinates perpendicular to the $z$-aligned guide field.}
              \label{fig:DW_diagram}%
    \end{figure*}
    
In fusion devices, drift-wave instabilities are the dominant source of anomalous (turbulent) transport \cite{Terry985, Rogister1996} and have thus been widely researched in fusion physics. These waves are formed in regions with a gradient in density and/ or temperature perpendicular to the strong background magnetic field, and where diamagnetic currents are induced to satisfy force balance. In this study, we will focus on density-gradient-driven DWs (as we expect that to be the dominant DWs mechanism in the solar corona due to the strong gradients existing across coronal loops), though other forms of DWs, such as ion- and electron-temperature-gradient-driven instabilities \cite{Lee1988}, trapped-electron modes \cite{Coppi1974} and micro-tearing modes \cite{Hazeltine1975} also exist. 

Density-gradient-driven DWs are generated in plasmas in which a density gradient $\nabla n$ exists perpendicular to the background magnetic field $\mathbf{B}_0$, see Figure~\ref{fig:DW_diagram}. If there is a perturbation $\tilde{n}$ to the number density profile such that locally, $n = n_0 + \tilde{n}$, the faster electrons will, on a short time scale, move either away or towards this region depending on whether $\tilde{n}$ is positive or negative, respectively. This will create regions of either positive or negative electric potential, $\tilde{\phi} (\tilde{n} > 0) > 0$ or $\tilde{\phi} (\tilde{n} < 0) < 0$. From the perturbed potential, an electric field ${\mathbf{\tilde{E}}} = - \nabla \tilde{\phi}$ is generated. Since $\mathbf{\tilde{E}}$ has a component perpendicular to $\mathbf{B}_0$, an ${\mathbf{E}} \times \mathbf{B}$ drift causes advection perpendicular to $\mathbf{B}_0$, hence the name drift wave. 

If the response of the electrons is adiabatic and $\tilde{n}$ and $\tilde{\phi}$ are in phase (the adiabatic response of the electrons is assumed to follow $\tilde{n} / n_0 = \tilde{\phi} / T_0$), neutral DWs are created. On the other hand, if the response is non-adiabatic, with a phase difference between $\tilde{n}$ and $\tilde{\phi}$, these waves can become unstable when the advection motion further amplifies the initial perturbation. This non-adiabatic response can, for example, stem from electron inertia or from the fact that there is collisionality in the plasma. In regions of higher density, there will be more collisions, and as a result, the electron response will be slowed down. In regions of lower density, the response will be faster in comparison, and thus, this will result in an accumulation of electrons in the regions where the density is higher already, amplifying the initial perturbation and destabilising the DW.


Once the unstable DWs have attained an adequate wave amplitude, stochastic heating sets in when the particle motion in an electrostatic wave becomes sufficiently chaotic that it can traverse significant regions of phase space. This heating has been explained theoretically as well as observed experimentally and numerically in nuclear fusion experiments \cite{McChesney1991} and the Earth's bow shock \cite{Stasiewicz2020}. Stochastic heating due to drift waves results in heating in the perpendicular direction, after which (partial) isotropisation may take place. Considering that the solar coronal heating problem has not yet been resolved {(see, e.g., the review of Klimchuk in Ref. \onlinecite{Klimchuk2006})}, this drift-wave-based heating is a candidate for further study. 

The observed perpendicular temperatures $T_\perp$ in the solar wind are indeed larger than the parallel temperatures $T_{||}$ (see, e.g., Ref. \onlinecite{Huang_2020} using Parker Solar Probe data). For that reason, it has been suggested by Vranjes and Poedts in Ref. \onlinecite{Vranjes2009a} that stochastic heating due to DWs may be at least partly responsible for coronal heating (for the entire rationale including approximate calculations of the predicted heating and signature of these waves, see Refs. \onlinecite{Vranjes2009, Vranjes2009a, Vranjes2009b, Vranjes2010c, Vranjes2010b, Vranjes2010, Vranjes2014}). Other arguments used in Ref. \onlinecite{Vranjes2009} which work in favour of this hypothesis are that stochastic heating is expected to heat protons indeed more efficiently than electrons, resulting in $\frac{T_{\text{e}}}{T_{\text{i}}} < 1$ (which is also observed in the solar wind). Moreover, heavier ions are heated more than lighter ions, which is also observed in the solar wind. 

Indeed, recent state-of-the-art EUV {(extreme ultraviolet)}, X-ray and even white-light observations have confirmed that the solar corona is highly filamentary and inhomogeneous (see, for example, Refs. \onlinecite{ Woo2007, Mackay2010, DeForest2018, Chitta2022, Antolin2023}), and thus, density gradients are ubiquitous on various scales and with a large range of magnitudes. For example, TRACE {(Transition Region And Coronal Explorer)} observations of coronal loops smaller than $1000\;$km have revealed the existence of structures with a thickness on the order of the spatial resolution of TRACE, see for instance, Ref. \onlinecite{McEwan2006} (the $0.5$'' resolution of TRACE corresponds to roughly $300\;$km on the Sun). Doppler measurements suggest that such fine structures may exist even at smaller scales of the order of $10\;$km \cite{woo2006}. These findings indicate that the solar corona provides an environment conducive to the formation of DWs and ensuant stochastic particle heating. 

Another mechanism through which heating can take place is kinetic particle acceleration if the DW mechanism leads to current sheet formation, magnetic reconnection and parallel electric fields. Particle heating due to magnetic reconnection on kinetic scales was, for example, studied by Pueschel et al. in Ref. \onlinecite{Pueschel2014} (though in this case, the initial setup imposed current sheets directly, and the DW mechanism was not considered as a driver thereof). 

Despite all these arguments, the possible presence of DWs in the solar corona has never been observationally studied. Numerical analyses were generally carried out through simplified numerical models assuming approximate plasma dispersion relations based on various approximations. A more sophisticated approach is necessary to determine at which frequencies and wavelengths these waves may be found in the corona and to assess quantitatively how their growth depends on the background plasma parameters. 
    
To that end, in the present study, we aim to characterise DWs under solar coronal conditions through linear gyrokinetic simulations. Once the linear behaviour of these DWs is understood, nonlinear simulations can be carried out in future work to evaluate turbulent amplitudes and heating rates.  

In Section~\ref{sec:methodology}, we introduce the code used to perform simulations and describe the numerical setup in detail. The results of the modelling are presented in Section~\ref{sec:results}. First, we show predictions for the DW frequencies and growth rates for the default parameter point at a variety of perpendicular wave numbers as well as for other varying background conditions such as a range of density gradients, electron $\beta$ and magnetic shear. This is  followed up by a comparison with drift-kinetic predictions. {Other possible coronal heating mechanisms are also discussed, and the expected DW characteristics are briefly compared to the properties of the other types of waves known to exist in the solar corona.} Finally, the physical interpretation of these results is provided including a discussion of the limitations of the study. The paper is concluded in Section~\ref{sec:conclusions}.

\section{Gyrokinetics with \textsc{Gene} and simulation setup}
\label{sec:methodology}

   \begin{figure*}
   \centering
   \includegraphics[width=10cm]{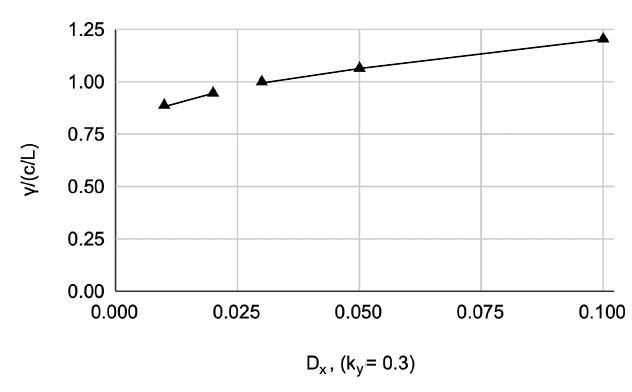}
   \caption{The non-dimensional growth rate as a function of hyper-diffusion $D_x$ in the $x-$direction for DWs at wavenumber $k_y = 0.3$.}
              \label{fig:hyperdiffusion}%
    \end{figure*}
    
The results presented in the present paper were obtained with the gyrokinetic Vlasov-Maxwell solver \textsc{Gene} \cite{Jenko2000} ({\url{https://genecode.org}). While this solver was originally designed for fusion research, there have been a number of studies in which \textsc{Gene} was used to simulate basic and astrophysical plasmas, including processes in the Solar corona \cite{Pueschel2011, Pueschel2014, Pueschel2015, Told2015, Told2016}.



The two main assumptions on which the gyrokinetic framework is based when compared to the regular kinetics are that it is assumed that i)~the background magnetic field variation happens on a much larger spatial scale than the Larmor radius, and ii)~the investigated frequencies are much lower than the ion gyrofrequencies, which in the corona lie in the range of $1$ to $100\;$kHz. These assumptions, while preventing the study of, e.g., light and cyclotron waves, allow for eliminating one velocity space coordinate due to gyro-averaging. Combined with the less restrictive time step due to ordering out the cyclotron motion, this results in a speed-up of simulations by multiple orders of magnitude. An assessment of the accuracy of gyrokinetics, compared to kinetic simulations, shows a good agreement between the two approaches even outside of the regime of the formal applicability of the former \cite{TenBarge2014}. For more detail on the theory and assumptions underlying gyrokinetics, we refer the reader to the reviews in Ref.  \onlinecite{Brizard2007} and Ref. \onlinecite{Garbet2010}.

The specific form and normalisation used in \textsc{Gene} is detailed by Pueschel et al. in Ref. \onlinecite{Pueschel2011}. For the presented case, a sheared slab was chosen as the magnetic geometry. {The particle distribution function consists of a background components and a fluctuating component. The initial condition for the fluctuating component of the distribution function $g_1$ is Maxwellian. The background distribution, $g_0$, remains Maxwellian throughout the run, but the fluctuating component $g_1$ is allowed to evolve. At the time of convergence of the eigenfunction, $g_0$ will in general not be Maxwellian.}

The reference length in the $z-$direction, $L_z$, can be identified as the loop length and $c_{\mathrm{s}}$ the ion speed of sound. The perpendicular wavenumber $k_y$ is normalised via the inverse Larmor radius $1/\rho_{\mathrm{s}}$. The frequency $\omega$ and growth rate $\gamma$ are normalised using $c_{\mathrm{s}} / L_z$. It should be noted that in our convention, positive frequencies correspond to drifts in the ion diamagnetic direction. The specific values corresponding to the conditions of the coronal loops will be inserted when discussing the physical interpretation of the results in Section~\ref{sec:results}. 

The discretisation of the domain was determined with a number of convergence tests performed on a range of different domain parameters. These included the number of grid points in the different directions considered, $N_{x}, N_{z}, N_{v_{||}}, N_{\mu}$, the box sizes of the extension of the velocity domain $L_{v_{||}}, L_{\mu}$, and the strength of hyper-diffusion (see Ref. \onlinecite{PueschelThesis} for its definition and rationale). Note that the box size $L_x$ is set by the parallel boundary condition and depends on $k_y$ and shear. The parameters of magnetic shear $\tilde{s}$, density gradient $\omega_n = \frac{L_z}{L_n}$ (where $L_n$ is the density gradient scale length), electron $\beta$ and the $k_y$ wavenumber were varied in this study. 

For the number of grid points $N_{x} \times N_{z}$ in the radial and parallel directions, $51\times 40$ was found to be generally sufficient, though in some cases investigated (especially with very low magnetic shear), $N_{x}$ of up to $131$ was needed at lower $k_y$'s. Next, the number of grid points in the parallel velocity direction $N_{v_{||}}$ and in the magnetic moment $N_{\mu}$ were found to be converged at $196$ and $16$, respectively. Converged velocity space domains are $L_{v_{||}} = 3$ and $L_{\mu} = 9$ in units of the species' velocity.

The fourth-order hyper-diffusion strength in the $x-$direction $D_x$ was set to $0.02$ based on the scan in Figure~\ref{fig:hyperdiffusion}, with only a moderate impact seen when varying this parameter. Note that $D_x > 0$ is required to ensure convergence at sensible values of $N_x$ due to its limiting role in mode extent in ballooning space \cite{Candy2004}. {Also note that in this and the following figures, the data points are notionally connected according to the branches that they are expected to belong to, as determined from the behaviour of their frequencies. {Mode-branch identification is based on continuous scaling in the frequency as well as smooth changes in the shape of the eigenfunction.}}

In the parallel direction, the fourth-order hyper-diffusion $D_z$ was set to $8$ with the precise value not affecting the results significantly over the range $1$ to $32$. The default set of physical input parameters reads $\omega_n = 100$, $\hat{s} = 0.25$ and $\beta = 0$.

\section{Drift-wave stability under coronal conditions}
\label{sec:results}



All simulations presented below are unstable to a DW arising from a prescribed background density gradient. An example drift-wave eigenfunction is given in Figure~\ref{fig:DW_diagram} on the right. The eigenfunction is shown in terms of the electrostatic potential $\phi$ and the  electron number density $n$. Note that the scale of these values is arbitrary as in this phase, the linear instability grows exponentially in time. However, relative amplitudes are meaningful. As mentioned above, it is clear that $\tilde{\phi}$ and $\tilde{n}$ are not in phase, a prerequisite for the instability developing in the first place. The spatial scale of these instabilities is expressed on the $x$ and $y$ axes in terms of the Larmor radii.


This drift wave behaves differently depending on the wavenumber and on the background plasma conditions, such as the density gradient, the magnetic shear, the electron $\beta$ and the presence of additional gradients, which can all influence how fast the DW  grows and at which frequencies it will drift. These dependencies are investigated in the sections below.

\subsection{Wavenumber dependency}


   \begin{figure*}
   \centering
   \includegraphics[width=\textwidth]{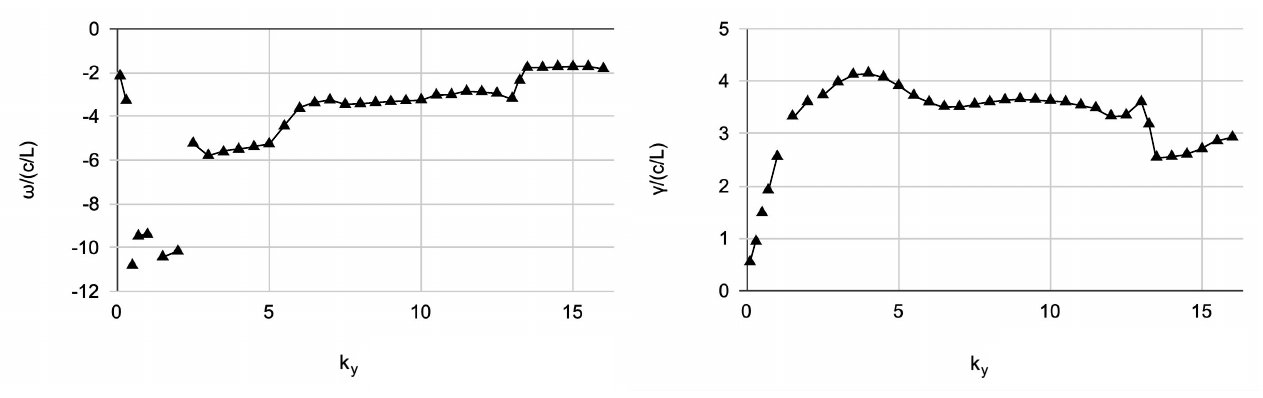}
   \caption{The dependency of DW real frequency (left) and growth rate (right) on $k_y$ for the base case.}
              \label{fig:Ky_with_branches}%
    \end{figure*}

First, it is evaluated how the frequency and the growth rates of the DW change for a varying perpendicular wavenumber $k_y$ for the base case. For $\hat{s} = 0.25$, $\omega_n = 100$ and $\beta = 0$, this behaviour is shown in Figure~\ref{fig:Ky_with_branches} for the frequency and for the growth rate. From this figure, it is evident that the non-dimensional frequencies range between 2.2 and 10.8, with the peak frequency at $k_y= 0.3$. The jumps in the frequencies between $k_y$ of 0.1 and 3 are due to the fact that there are several DWs present with similarly large growth rates but with different eigenfunctions (corresponding to different characteristic $k_z$), and each time, it is a different mode that ends up dominating the growth. {Indeed, different branches may arise from the same dispersion relation that respond differently to an increasing $k_y$ and parameters such as $\hat{s}$ and $\beta$. This may then result in the fact that these branches have similar growth rates in certain conditions. For more details, consult, for instance, Ref. \cite{Pueschel2008}.} Figure~\ref{fig:Ky_with_branches} shows that the growth rates vary between 0.6 and 4.6, with the peak growth rates occurring at $k_y = 4$. 

Since the peak frequency occurs at $k_y = 0.3$ and the peak growth rate at $k_y = 4$, it is these two non-dimensional wave numbers that will be mostly investigated from here on. Also note that based on mixing-length arguments, the expected nonlinear (turbulent) spectral peak lies at possibly substantially lower $k_y$ than where the fastest growth occurs.

\subsection{Density gradient dependency}

   \begin{figure*}
   \centering
   \includegraphics[width=\textwidth]{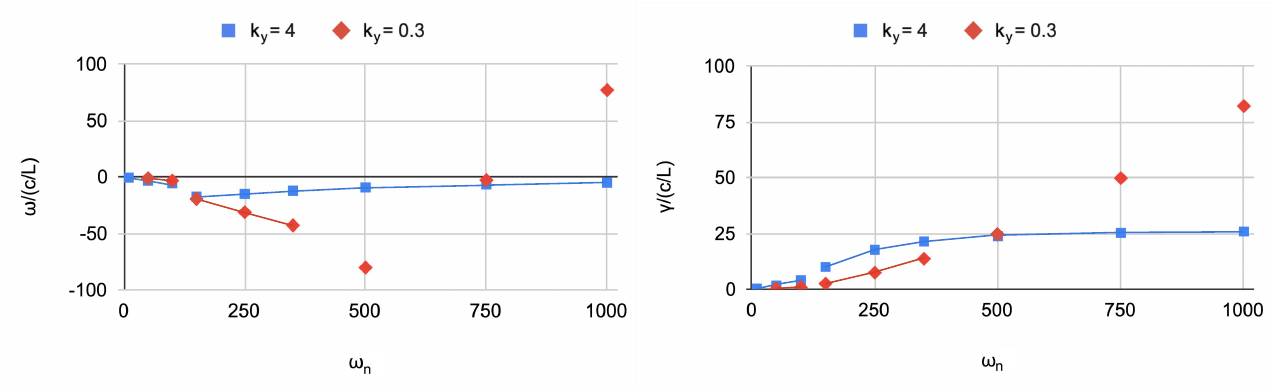}
   \caption{The dependency of DW real frequency (left) and growth rate (right) on the density gradient $\omega_n$.}
              \label{fig:GradientDensity_norho_nofinal}%
    \end{figure*}
    
    
The instability drive of the DW is the density gradient. The baseline simulations presented above were computed using a nondimensional density gradient of 100, where this gradient is defined as,

\begin{equation}
    \omega_n = -\frac{L_{\text{z}}}{n_0} \frac{dn_0}{dx}.
\label{eq:densitygrad}
\end{equation}

Here, $n_0$ is the number density and $dn_0/dx$ the density gradient which defines the $x-$direction. In the solar corona, density gradients can be assumed to vary substantially. For instance, Cargill and coworkers\cite{Cargill2016} studied the evolution of density gradients in coronal loops through zero-dimensional modelling and found that $dn_0/dx$ $\sim$ $5\;$cm$^{-4}$ to $20\;$cm$^{-4}$ for their specific cases. Using these values for $dn_0/dx$ along with a reference parallel length of $L_{\text{z}}$ of $10^8\;$m (to represent an average coronal loop length) and a number density of $10^{14}\;$m$^{-3}$ as assumed in Ref. \onlinecite{Cargill2016}, we obtain $\omega_n$ = $100$ to $1000$. This justifies our initial choice for $\omega_n$, and also suggests that higher values may be relevant to the corona. In addition, in order to include weaker coronal density structures, we investigate $\omega_n$ {down to the value of 1}. 

The behaviour of the DW frequencies and growth rates depending on the prescribed density gradient is shown in Figure~\ref{fig:GradientDensity_norho_nofinal}. {Since there was no instability occurring at $\omega_n = 1$ for both wavenumbers and for $\omega_n = 10$ for $k_y = 0.3$, these points are omitted}. As expected, the growth rate increases with $\omega_n$, in our case, by two orders of magnitude between $\omega_n=10$ and $\omega_n=1000$. There are also significant changes in the resulting frequencies, demonstrating that the frequency of the DW generated by a weak gradient can be an order of magnitude smaller compared to the DW coming from a region with stronger gradients. {We also see that at very large density gradients, the frequency flips its sign, meaning that the drift changes its direction, from the electron diamagnetic direction to the ion diamagnetic direction. Such reversals are not uncommon in drift-wave turbulence\cite{Faber2015}.}


{The critical $\omega_{n, \text{crit}}$ (i.e., an $\omega_n$ which has to be exceeded in order for the instability to occur) for $k_y = 0.3$ was observed to be between $10 < \omega_{n, \text{crit}} < 50$, whereas for $k_y = 4$, $\omega_{n, \text{crit}} \approx 1$. In DW turbulence, indeed, instability thresholds are known to depend on the wavenumber\cite{Pueschel2008}. The value of the critical gradient} may change, however, if effects such as field-line curvature are introduced. Furthermore, this finding does not preclude the existence of a (substantial) nonlinear critical gradient that may arise from nonlinear resonances \cite{Terry2021, Pueschel2021Resonance, Li2021, Li_2023}. For $k_y = 0.3$, the instability no longer develops for $\omega_n < 20$.

\subsection{Magnetic shear dependency}

  \begin{figure*}
   \centering
   \includegraphics[width=\textwidth]{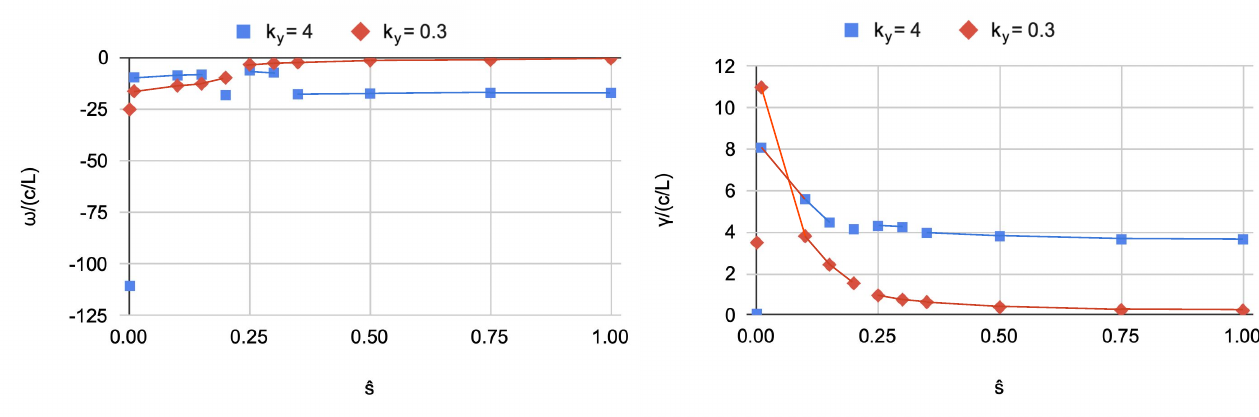}
   \caption{The dependency of the frequency (left) and the growth rate (right) on the magnetic shear $\hat{s}$. Note that the smallest shown value in this plot corresponds to $\hat{s} = 0.001$ as $\hat{s} = 0$ would not result in an instability.}
              \label{fig:Shat_final_norho}%
    \end{figure*}

Another aspect affecting the behaviour of the DW is the magnetic shear. In fusion devices, the magnetic shear $\hat{s}$ is determined from the profile of the device safety factor as

\begin{equation}
    \hat{s} = \frac{r_{0}}{q_{0}} \frac{dq(x)}{dx},
\end{equation}

where the safety factor profile $q(x)$ is taken near a flux surface located at a distance $r_0$ (corresponding to $q_0$) from the magnetic axis. Note that $q_0$ corresponds to the twist of the magnetic field lines. 

{There is very little little knowledge} about the field-line-twist profiles of coronal loops, however. Aschwanden in Ref. \onlinecite{Aschwanden2019} found that the average $N_{\text{twist}}$ (which can be interpreted as $\approx 1/q_0$) in coronal loops is 0.14, giving a $q_0$ of 3.6 (assuming a half loop). However, to evaluate $\hat{s}$, the profile of $q(x)$ at the flux surface is needed, which, to the authors' knowledge, has not yet been determined for coronal loops. If we assume the extreme case that $q$ is 0 in the middle of the loop, it is possible to compute the maximum $\hat{s}$. In Ref. \onlinecite{Aschwanden2019}, Aschwanden suggests a typical loop inverse aspect ratio of $r_{\text{loop}}/L_{\text{loop}} \approx 0.1 - 0.4$, so $0.1L_{\text{z}}$ can be seen as a characteristic flux surface minor radius at which $q = 3.6$. In that case, assuming we have the given $q_{\text{max}}$ on that surface and $q = 0$ in the middle, we arrive at a maximum estimate for the magnetic shear of $\hat{s}_{\text{max}} \approx 1$.

In addition, it is expected that the magnitude of magnetic shear in the large variety of density structures in the solar corona is similarly variable. For that reason, below, we investigate the DW growth rates for an entire spectrum of magnetic shear, ranging from very small values {of $\hat{s} = 0.001$ and $\hat{s}=0.01$ up to $\hat{s} = 1$. N}ote that finite $\hat{s}$ is required for DW to be unstable.

Figure~\ref{fig:Shat_final_norho} shows the results for the frequency and growth rate at $k_y  = 0.3$ and 4, respectively, as functions of the shear $\hat{s}${, with the smallest value shown being $\hat{s} = 0.001$}. As expected, a very low shear causes the growth rates to drop, whereas for a large $\hat{s} \gtrsim 0.5$, a plateau develops. Apart from the cases with very small shear, however, most of the complex eigenfunctions lie in the general vicinity of the result for the default case $\hat{s} = 0.25$. 


  \begin{figure*}
   \centering
   \includegraphics[width=\textwidth]{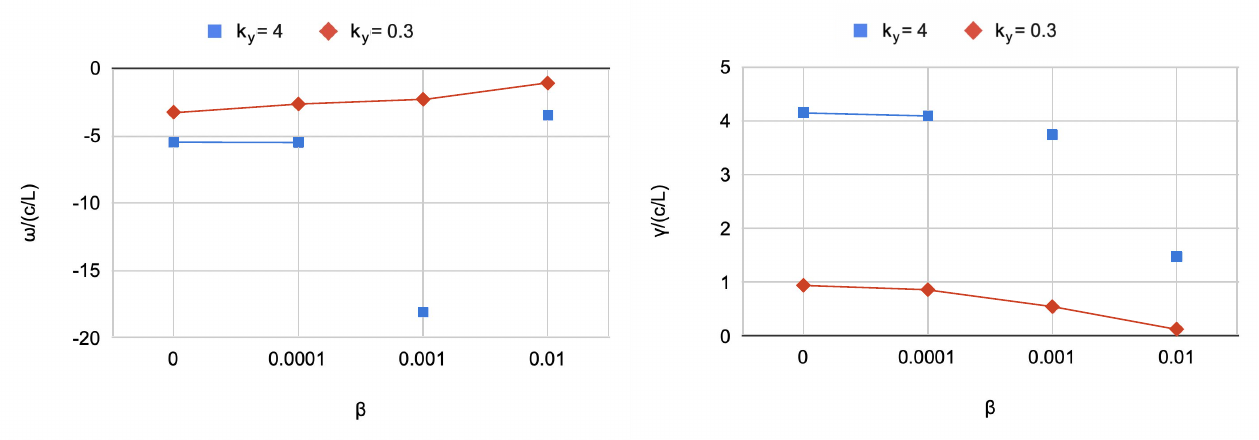}
   \caption{The dependency of the frequency (left) and the growth rate (right) on the electron $\beta$.}
              \label{fig:Beta_final_norho}%
    \end{figure*}


\subsection{Electron $\beta$ dependency}

So far, it has been assumed that electron $\beta$ (the ratio of the electron thermal to the magnetic pressure) is 0 for simplicity, and thus, we are considering a very strongly magnetised plasma, which is favourable for the excitement of the drift wave. However, according to Ref. \onlinecite{Gary2001}, even in the magnetically more active regions in the lower corona, $\beta$ is expected to vary between roughly $0.0001$ and $0.01$. For that reason, we computed the frequencies and growth rates for finite $\beta$ to determine how the formation of magnetic fluctuations - the consequence of $\beta > 0$ - and the concomitant alteration of the confining field affect DW characteristics. 

Figure~\ref{fig:Beta_final_norho} presents the variation of the DW frequency and growth rate over a range of $\beta$'s for $k_y=0.3$ (red diamonds) and $4$ (blue squares). The remaining physical input of the parameters corresponded to the base case, $\hat{s} = 0.25$ and $\omega_n = 100$. 

As the $\beta$ increases, the growth rate decreases significantly. However, even at the highest tested $\beta$, this DW growth rate is still positive. We observe a drop by a factor of roughly $10$ over two orders of magnitude increase in $\beta$ for $k_y=0.3$, and only a factor of $3$ for $k_y=4$. This indicates that DWs may also be unstable in magnetically less active regions, especially if the corresponding density gradients are to compensate. We also see that with the exception of one data point, the frequencies can be expected to decrease with increasing $\beta$. The $k_y = 4$ data point at $\beta = 10^{-3}$ shows that there are again multiple strong-growing modes with different frequencies present, and the identity of the linearly dominant mode branch can sensitively depend on the specific plasma conditions \cite{Pueschel2019PPCF}.

\subsection{DWs in the presence of temperature gradients}

The solar corona is also characterised by temperature gradients, though these can be expected to be generally milder than the density gradients (e.g., across the cross-section of a coronal loop). Temperature-gradient-driven modes can coexist with density-gradient-driven modes and DW properties can change based on a mixed gradient drive. In regions such as prominences with locally cool, condensed plasma (see, e.g., the simulations carried out by Jenkins et al. and Brughmans et al. in Ref. \onlinecite{Jenkins2021} and Ref. \onlinecite{Brughmans2022}), these effects may be especially important.  

For that reason, the base case with $\omega_n = 100$ was modified with the inclusion of a temperature gradient $\omega_T = 1, 10, 50$ and 100, where the temperature gradient is defined analogously to the density gradient.

The results are shown in Figure~\ref{fig:GradT_final_norho} for the frequency and for the growth rate. In both plots, the blue squares correspond to $k_y = 4$ and the red diamonds to $k_y = 0.3$.

  \begin{figure*}
   \centering
   \includegraphics[width=\textwidth]{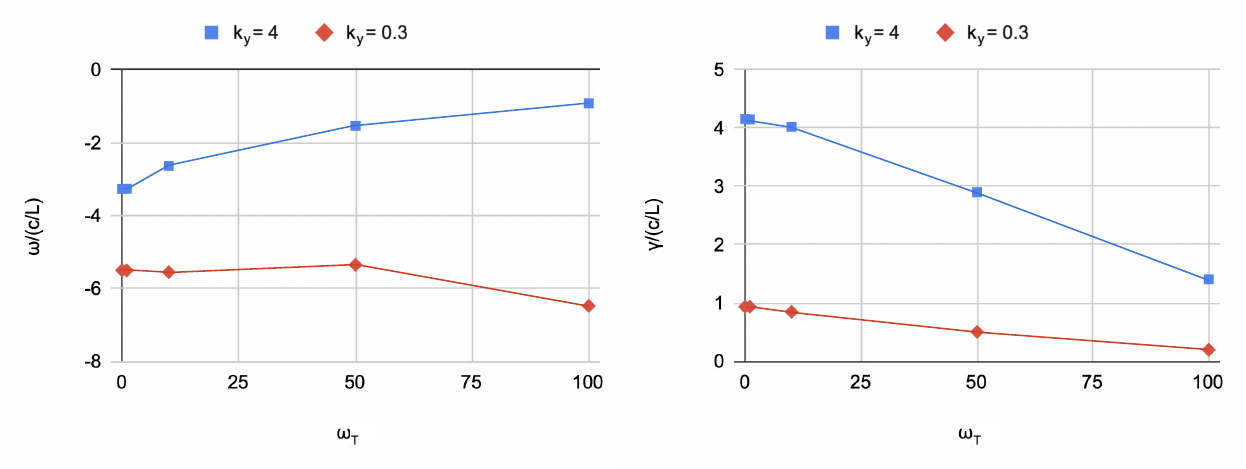}
   \caption{The dependency of the frequency (left) and the growth rate (right) on the added temperature gradient $\omega_T$.}
              \label{fig:GradT_final_norho}%
    \end{figure*}


{For small temperature gradients $\omega_T \leq 10$}, the resulting instability is not significantly affected for the investigated wave numbers (with the frequencies and growth rates remaining within $10\%$ of the base case). At larger temperature gradients, the DWs become partly stabilised. Mathematically, this property arises from the drive term $\propto \omega_n + (v_\parallel^2 + \mu B_0 - 3/2)\omega_T$ in the Vlasov equation \cite{Pueschel2011}, where for $v_\parallel^2 < 3/2 - \mu B_0$, the temperature gradient can act in a fashion opposing the density gradient. However, as long as $\omega_n \gg \omega_T$, one can expect DWs to remain strongly unstable. 

\subsection{Comparison with drift-kinetic simulations}

Finally, we have seen above that there is a significant growth at relatively large perpendicular wave numbers $k_y > 1$. We now investigate whether DWs are affected by finite Larmor radius (FLR) effects. This can be determined by comparison with drift-kinetic simulations, which, in addition to the gyrokinetic approximations, assume that the particle gyroradius is small compared to the fluctuating scales, $k_y \ll 1$. In practice, this means that Bessel functions of argument $k_y$ in the gyrokinetic equations are replaced by a truncated Taylor series. 

  \begin{figure*}
   \centering
   \includegraphics[width=12cm]{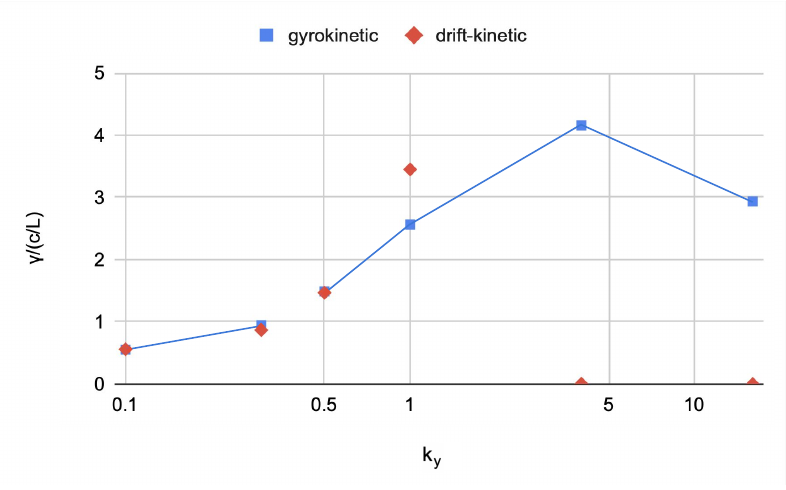}
   \caption{Comparison of growth rates as functions of perpendicular wavenumber as computed according to gyrokinetics (blue squares) and drift-kinetics (red diamonds). Note that the two gyrokinetic and drift-kinetic data points overlap at $k_y = 0.1$.}
        \label{fig:Gyro_vs_drift_final_norho}%
    \end{figure*}

The resulting comparison can be seen in Figure~\ref{fig:Gyro_vs_drift_final_norho}, where the gyrokinetic growth rates are shown as blue squares, while the drift-kinetic growth rates are shown as red diamonds.

At lower $k_y$, the predicted growth rates according to gyrokinetics and drift-kinetics agree well with each other{, with the values at $k_y = 0.1$ matching closely}. At higher wave numbers, however, the drift-kinetic predictions for the growth fall to zero. This shows that drift-kinetics is inaccurate at gyroradius and sub-gyroradius scales ($k_y \gtrsim 1$), meaning that  FLR effects play a significant role in the destabilisation of small-scale DWs.



\subsection{Application to the solar corona}

To understand what the findings above imply for the solar corona, in this section, the results will be translated into dimensional quantities. 

The baseline reference values were chosen as follows. The reference length in the $z$-direction $L_{\text{z}}$ of $100\;$Mm was used to represent the length of a typical coronal loop. The background parallel magnetic field $B_{\text{0}}$ (also in the $z-$direction) was set to $10\;$G, or $0.001\;$T, to represent the typical magnetic field strength in coronal loops. The background temperature $T_{\text{0}}$ was set to $90\;$eV, or roughly $1\;$MK, also representative of the electron (and ion) temperature in coronal loops. For typical ranges of values of these quantities in coronal loops, see the work of Dahlburg et al. in Ref. \onlinecite{Dahlburg2018}. 

Using the reference speed of sound of the ions $c_{\mathrm{s}}$, as derived from the reference temperature and the reference length, we obtain a normalisation frequency $\omega_{\text{ref}} =  c_{\mathrm{s}} / L_z$ of $\approx 9 \cdot 10^{-4}\;$rad s$^{-1}$ = $1.4 \cdot 10^{-4}\;$Hz. This means that for our base case, the non-dimensional frequencies (in the range of $2$ to $12$) and the non-dimensional growth rates (in the range of $0.5$ to $5$) correspond to real frequencies of $0.3\;$mHz to $1.7\;$mHz and growth rates of $0.07\;$mHz to $0.7\;$mHz. For smaller structures (e.g., shorter coronal loops), we find that the implied real frequencies and growth rates increase inversely in proportion, that is, of $3\;$mHz to $17\;$mHz frequencies for $1\;$Mm structures and $3\;$Hz to $17\;$Hz for $10\;$km structures. The same scaling applies to the growth rate. 

The background temperature affects the reference speed of sound, and thus, colder or hotter structures will also have different ranges of frequencies and growth rates. Considering the base case, a lower temperature of $100\;$kK results in frequencies of $0.09\;$mHz to $0.6\;$mHz and growth rates of $0.02\;$mHz to $0.2\;$mHz. A much hotter coronal loop of $10\;$MK, in contrast, would yield frequencies between $0.9\;$mHz to $6\;$mHz. 

The ranges above only hold for the base case, the results of which were presented in Figure~\ref{fig:Ky_with_branches}. If we significantly either increase or decrease the density gradient, we correspondingly get much higher or smaller values for the growth rate, respectively. For the investigated case of $k_y = 4$, the maximum growth rate of $0.6\;$mHz increases to $4\;$mHz at $\omega_n$ = $1000$, and gets reduced down to $0.04\;$mHz at $\omega_n$ = $10$. For a varying $\beta$, wave frequencies remain in the $0.1\;$mHz to $2\;$mHz range, with the growth rate falling down to $0.02\;$mHz at $k_y = 0.3$ for the highest $\beta$ tested.

   \begin{figure*}
   \centering
   \includegraphics[width=\textwidth]{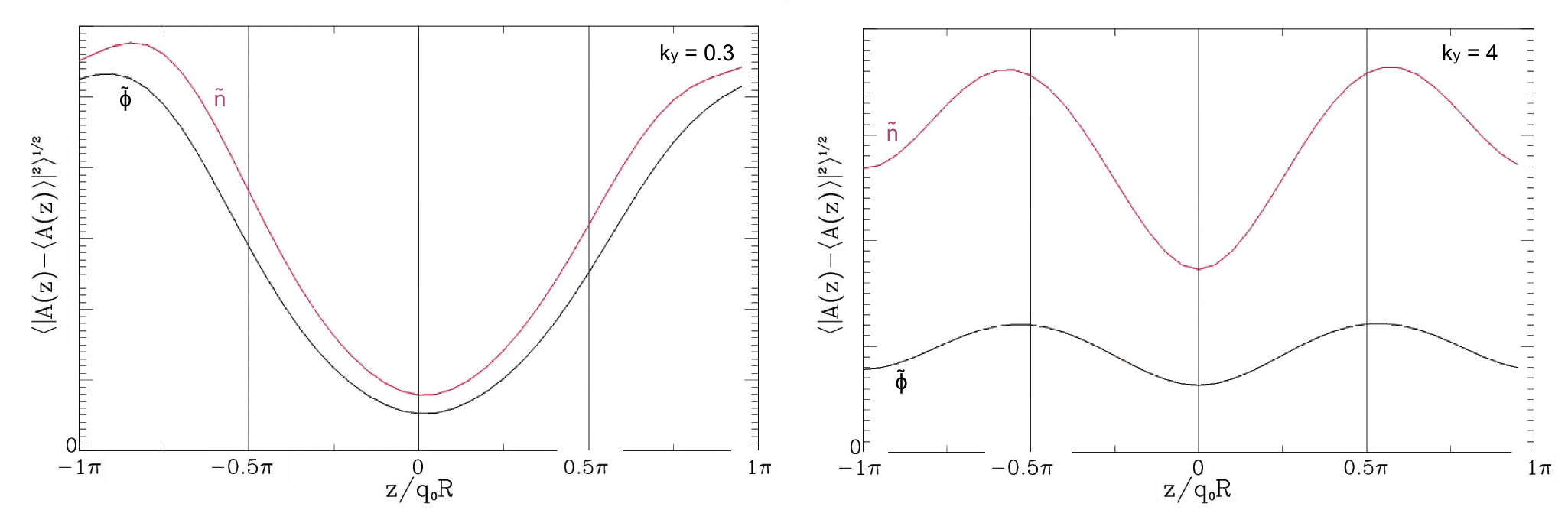}
   \caption{The amplitude fluctuations of the electrostatic potential and density in the direction along the coronal loop path ($z$-axis). The values on the vertical axes are arbitrary.}
              \label{fig:potential_zprofile}%
    \end{figure*}

Finally, the typical ion gyroradius for our conditions, considering a background coronal magnetic field of $10\;$G and the sound speed of ions of $90\;$km/s, is roughly $1\;$m. Thus, the indicated wave numbers at which these waves grow fastest and have the largest frequencies also correspond to $k_y \approx 0.3$ to $5\;$m$^{-1}$, which translates into wavelengths of $\approx 1$ to $20\;$m. These wavelengths are much smaller than any scale present-day space instruments can resolve in the corona. DWs also propagate in the direction parallel to the magnetic field, with $k_{||} \ll k_\perp$ (see observations in Ref. \onlinecite{McChesney1991}). This direction is along the loop, and our simulations indicate that the first and second harmonics in $k_{||}$ are the dominant mode, see Figure~\ref{fig:potential_zprofile}, depending on the choice of $k_y$. That is, the wavelength of this oscillation is half the loop radius for $k_y = 4$ or the full loop radius for $k_y = 0.3$. However, linear results by themselves are insufficient to predict the amplitudes of these oscillations and, thus, whether they should be observable, especially next to the strong modes such as those generated by the kink instabilities.

The frequency values indicated above are interesting from the observational standpoint. Kepko et al. in Ref. \onlinecite{Kepko2003} identified density fluctuations with frequencies between $0.1$ and $3\;$mHz in the solar wind plasma. In Ref. \onlinecite{Kepko2003} and Ref. \onlinecite{Kepko2002}, it is suggested that these periodic density fluctuations may originate in the solar corona, which was confirmed in by Viall et al. in Ref. \onlinecite{Viall2009} and Ref. \onlinecite{Viall2021Periodic}. The exact driver of these fluctuations remains unknown to this day. If these density fluctuations are accompanied by fluctuations in the electrostatic potential, DW could be a viable candidate for their explanation. {This discussion thus brings us to the point where we can compare the DW properties determined above to the properties of other mechanisms that suspected to contribute towards coronal heating.}

\subsection{Comparison with other heating mechanisms}
\label{subsec:comparisonother}

{In this work, we treat DWs as one of the potential candidates to (at least partly) explain the source of solar coronal heating. There has been, however, a variety of other mechanisms suggested for this role, and it is thus important to compare DWs to these mechanisms to determine in which aspects they may differ in practice and how they may be distinguished.}

{The mechanisms that could be responsible for coronal heating are typically separated into the those involving dissipation of energy from waves and those related to magnetic reconnection, see, for instance, the review in Ref. \onlinecite{Klimchuk2006}. While originally, studies mostly focused on finding "the one" coronal heating mechanism, recently it has been more and more accepted by the community that there is not a single one mechanism solely responsible for heating the corona, but that more likely, the heating is caused by a variety of mechanisms at different spatial and temporal scales\cite{DeMoortel2015}. A review of the variety of waves that have been observed in the solar corona and that could contribute towards coronal heating can be found in, for example, the work of Van Doorsselaere et al. in Ref. \onlinecite{VanDoorsselaere2020}, including Alfvén waves, magnetosonic waves and standing waves. As has been also pointed out in Ref. \onlinecite{DeMoortel2015} however, it can be difficult to separate the different mechanisms, as MHD waves may drive magnetic reconnection and vice versa. Similarly, it was in Ref.~\onlinecite{Pueschel2014} that magnetic reconnection in coronal plasma may lead to the generation of kinetic Alfvénic waves. Alfvénic turbulence has been shown to be a viable candidate to explain coronal heating\cite{vanBallegooijen2011}.}

{The main feature which makes drift waves distinguishable from the other observable wave types, such as Alfvén waves, is the presence of electrostatic fields of
substantial amplitudes. In Figure~\ref{fig:DW_diagram} on the right, we see oscillations in the electrostatic potential $\tilde{\phi}$, which, provided that their amplitude (as determined from nonlinear simulations) is large enough, may be observable in future in-situ measurements.}

{The different kinds of wave mechanisms provide heating at different spatial and temporal scales. For example, for standing kink waves, frequencies between $0.6\;$mHz to $17\;$mHz are reported\cite{Nechaeva2019}. The CoMP waves, observed by the Coronal Multi-channel Polarimeter\cite{Tomczyk2008}, have been interpreted as transverse Alfvén waves\cite{Tomczyk2007, TomczykMcIntosh2009} with a peak at $\approx 3.2\;$mHz, though Van Doorsseleare et al. argue that these could be fast-mode kink waves instead\cite{VanDoorsselaere2020}. Khan et al. predict a few mHz for kinetic and inertial Alfvén waves in Ref. \onlinecite{Khan2020}. For decayless standing waves observed in coronal loops, the survey of which was performed in Ref. \onlinecite{Tian2012}, frequencies of 3 to 6 mHz were reported. We thus see that the frequencies of DWs expected from the results reported in the present paper overlap with these quite significantly.} 

{However, as also visible on Figure~\ref{fig:DW_diagram}, the spatial scale of the DW oscillations is very small, on the order of a few to a few tens of Larmor radii depending on the wavenumber. In solar coronal conditions, this means the scale of a few centimeters to a few meters. Given the high speeds (generally $\gg 1\;$kms$^{-1}$) of the CoMP and standing waves summarised in the work in Ref. \onlinecite{VanDoorsselaere2020}, the corresponding wavelengths would be mostly in the range of hundreds of kilometers to megameters. In the conditions of the solar corona, DWs would thus be observable on much smaller spatial scales. Further away into the heliosphere, however, thanks to the smaller magnetic field and thus much larger gyroradii, the wavelengths of DWs may also reach the scale of kilometers to megameters, however.}

{Finally, in the nonlinear simulations planned in the future, it will be also possible to analyse the resulting turbulence spectrum and compare that with what has been measured in-situ in the solar wind, similarly to what has been done by in Ref. \onlinecite{Pueschel2014}. These nonlinear simulations can be set up in the future thanks to the linear simulations presented in this paper.}

\subsection{Limitations of the modelling approach}
\label{sec:discussion}

The results above suggest the possible existence of unstable drift waves in coronal loops with frequencies and growth rates lying in the general range of $0.1\;$mHz to $1\;$Hz, which are unstable in all of the conditions tested and which grow especially in environments with high magnetisation and density gradients. It must be emphasised, however, that the ranges of frequencies and growth rates estimated for this mode were computed with linear simulations only. Thus, these simulations cannot provide an estimate of the oscillation amplitudes. {
In addition, quasilinear treatments (see, e.g., Ref.~\onlinecite{Dannert2005}) of drift-wave instabilities produce turbulent diffusion scaling as $\gamma / k_\perp^2$, allowing large scales to affect the turbulence much more substantially than small scales. Relatedly, nonlinear characteristics of drift-wave turbulence, such as spectra, cross-phases and probability distribution functions, are generally set by the larger scales in the system\cite{Jenko2005}, whereas coherence is lost and amplitudes are much smaller at the smaller scales, even when growth rates there remain substantial. See also Ref.~\onlinecite{Staebler2016}. In practice, this means} that nonlinear fluctuation amplitudes will peak at scales larger than that associated with the fastest-growing linear eigenmode. 

For these reasons, we have not yet proceeded with the estimation of DW-based heating of the coronal plasma. Corresponding nonlinear analyses are ongoing and will be reported at a later time.

In addition, it should be noted that here, only a simple slab geometry with a density gradient is studied. In reality, magnetic curvatures  may result in additional dynamics that affect the outcomes. For example, field-line curvature can result in critical $\omega_n$, below which no instability exists. 

{Further 3D effects include the fact that in solar wind, electrons largely do not have Maxwellian distributions, but contain suprathermal tails, with a distinguishable thermal core, an energetic halo and a more energetic superhalo\cite{Ko1996,Lin1998}. These may, according to numerical simulations, also exist in the solar corona, see, for instance, the work of Vocks et al. in Ref.~\onlinecite{Vocks2008} or Cranmer in Ref.~\onlinecite{Cranmer2014}, and they may have an influence on the behaviour of DWs according to the work of Guo in Ref.~\onlinecite{Guo2023}. In the work of Guo, it is shown that the critical wavenumber may become higher when the suprathermal features are included and that both the frequency and the growth rates may be affected by a few tens of percent. However, the Strahl component of the electron halo is, unlike the core and the superhalo, highly anisotropic, and generally points in the direction away from the Sun\cite{Lin1998,Vocks2008}. Thus, its direction is expected to change depending on the position along the coronal loop. Such considerations can inform future studies that extend the present effort to realistic, 3D coronal-loop geometries.}



\section{Conclusions and outlook}
\label{sec:conclusions}

In this paper, we use the gyrokinetic code \textsc{Gene} to compute linear properties of drift waves (DWs) in the environment of the solar corona. DWs can get excited linearly in magnetised plasmas that contain density and/or thermal gradients perpendicular to the magnetic field due to the different dynamics of ions and electrons, and a non-adiabatic response of electrons further destabilises these modes. {It was shown in Refs. \onlinecite{Vranjes2009, Vranjes2009a, Vranjes2009b, Vranjes2010, Vranjes2010b, Vranjes2010c, Vranjes2014}} that DWs may potentially contribute to coronal heating through processes such as particle acceleration and stochastic heating, which may explain phenomena such as temperature anisotropy of the solar wind and stronger heating of heavier ions. Especially in the solar coronal environment, the density-gradient-driven DW is relevant, considering the profiles of coronal loops.

Analysis of density-gradient-driven DWs  within the gyrokinetic framework, which, in comparison with kinetics, averages over the gyrating motion of the charged particles and thus requires fewer computational resources. The setup considered was a slab geometry with an ion-electron plasma with a background density gradient and a background magnetic field perpendicular to it.

Linear results indicate that for all investigated conditions, the drift-wave mechanism is present and unstable, except for smaller $k_y$'s at small density gradients. Simulations exhibit the growth rates in the range between $0.07$ and $0.7\;$mHz and frequencies in the range between $0.3$ and $1.7\;$mHz for our base case (a $1\;$MK corona with a $10\;$G magnetic field, a loop length of $100\;$Mm and a nondimensional density gradient of $100$, as derived from Ref. \onlinecite{Cargill2016}). Frequencies of up to tens of mHz to Hz are estimated for very hot ($> 10^7\;$K) and small ($< 10\;$km) density filaments. 

We also demonstrate that increasing the density gradient enhances the growth of this instability. While finite magnetic shear is required for growth, for realistic values, growth rates are rather insensitive to shear.  Finally, increasing electron $\beta$ reduces growth by roughly one order of magnitude going from the electrostatic limit, $\beta = 0$, to $\beta$ = $0.01$, which is more representative of the coronal environment. 

{This work thus concludes that DWs may be present and unstable in the environment of the solar corona. In comparison to the work of Vranjes and Poedts \cite{Vranjes2009, Vranjes2009a, Vranjes2009b, Vranjes2010, Vranjes2010b, Vranjes2010c, Vranjes2014}, we base this conclusion on an extensive numerical gyrokinetic study instead of a simplified dispersion relation, with this study carried out for a wide variety of wavenumbers, density gradient strengths, magnetic shear strengths, realistic $\beta$ values and even in the presence of additional temperature gradients. We also show that the predicted frequencies of these simulations agree with the periodic oscillations that have been recently measured in the solar wind \cite{Viall2021Periodic}. Finally, we compare the expected characteristics of DWs with the properties of the other types of waves present in the solar corona that are currently suspected to contribute towards coronal heating.} 



However, linear simulations are not sufficient to predict to which extent this mechanism can contribute to coronal heating.  They cannot determine the amplitudes of the resulting oscillations. Our linear results are merely indicative that DWs may be present and unstable in the corona, and they show in which environments this mode should be expected to be more prevalent and faster growing. Determining the linear characteristics is, however, an important first step towards carrying out a more sophisticated analysis. Knowing how these modes behave linearly allows us now to progress to nonlinear simulations, through which we can determine the heating as well as heat flux that can be generated via this mechanism. Nonlinear simulations will indicate on which time scales DWs can affect the background and how these time scales compare with a typical coronal-loop lifetime. Finally, comparisons with observations will be possible through the analysis of the resulting turbulence spectrum. This work is currently in progress.

\begin{acknowledgments}
This research was funded with projects C14/19/089  (C1 project Internal Funds KU Leuven), G.0B58.23N and G.0025.23N (FWO-Vlaanderen), 4000134474 (SIDC Data Exploitation, ESA Prodex-12), and Belspo project B2/191/P1/SWiM. The resources and services used in this work were provided by the VSC (Flemish Supercomputer Centre), funded by the Research Foundation - Flanders (FWO) and the Flemish Government, and by the Tier 1 VSC grant project 2023/019.
\end{acknowledgments}

\section*{Data Availability Statement}
The data that support the findings of
this study are available from the
corresponding author upon reasonable
request.

\bibliography{biblio.bib}

\end{document}